\documentclass[final]{aipproc}
\layoutstyle{8x11double}
\usepackage{epstopdf}

\usepackage{graphicx}

\usepackage[usenames]{color}
 \definecolor{magenta}{cmyk}{0,1,0,0}

 \definecolor{blue}{rgb}{0,0,1}

\newcommand{\onepion}{one pion\ }

\newcommand{\GeV}{\; \mathrm{GeV}}

\newcommand{\dd}{\mathrm{d}}

\begin{document}

\title{Comparison of GiBUU calculations\\ with MiniBooNE pion production data}
\author{O. Lalakulich}{
  address={Institut f\"ur Theoretische Physik, Universit\"at Giessen, Germany}
}
\author{U. Mosel}{}

\begin{abstract}
\begin{description}
\item[Background] Neutrino-induced pion production can give important informationon the
axial coupling to nucleon resonances. Furthermore, pion production represents a major background
to quasielastic-like events. \onepion production data from the MiniBooNE in charged current  neutrino scattering in
mineral oil appeared higher than expected within conventional theoretical approaches.
\item[Purpose] We aim to investigate which model parameters affect the calculated
cross section and how they do this.
\item[Method] The Giessen Boltzmann--Uehling--Uhlenbeck (GiBUU) model is used for an
investigation of neutrino-nucleus reactions.
\item[Results] Presented are integrated and differential cross sections for 1$\pi^+$ and 1$\pi^0$
production before and after final state interactions in comparison with the MiniBooNE data.
\item[Conclusions] For the MiniBooNE flux all processes (QE, 1$\pi$-background, $\Delta$,
higher resonance production, DIS) contribute to the observed final state with one pion
of a given charge.
The uncertainty in elementary pion production cross sections leads to a corresponding uncertainty in the nuclear cross sections.
Final state interactions
change the shape of the muon-related observables only slightly, but they significantly change
the shape of pion distributions.
\end{description}
\end{abstract}

\date{\today}

\maketitle

\section{Introduction}

At NuInt 2009 a broad comparison of generator predictions for QE and \onepion cross
sections~\cite{Boyd:2009zz} has shown major discrepancies
($\approx 30$\% for QE, up to 100\% for pion production)
between the predictions of various generators presently being used by neutrino experiments.
At this conference (see the contribution of P. Rodrigues), a similar comparison shows
more consistent results, because now each neutrino event generator can be fine tuned
to the recent MiniBooNE data on charged~\cite{AguilarArevalo:2010bm}
and neutral~\cite{AguilarArevalo:2010xt} pion production in charged current (CC) neutrino scattering.

Does the fine tuning reveal the physics behind it and contribute to our understanding
of the microscopic mechanism inside nucleus? Is this fine tuning done consistently for
pion production and QE scattering? Answering these questions requires that not only
the results (as have been done at this conference),
but also the physical models behind the generators are compared.
One would need to identify the crucial physical inputs responsible for the
key features of predicted cross sections and to use those inputs consistently.
We hope that a fruitful discussion between different generator-building groups is coming and
this discussion starts here.


\section{Input}

\subsection{Accuracy of elementary input determines the accuracy of calculated cross sections}

\begin{figure}
\begin{minipage}[c]{0.32\textwidth}
\includegraphics[width=\textwidth]{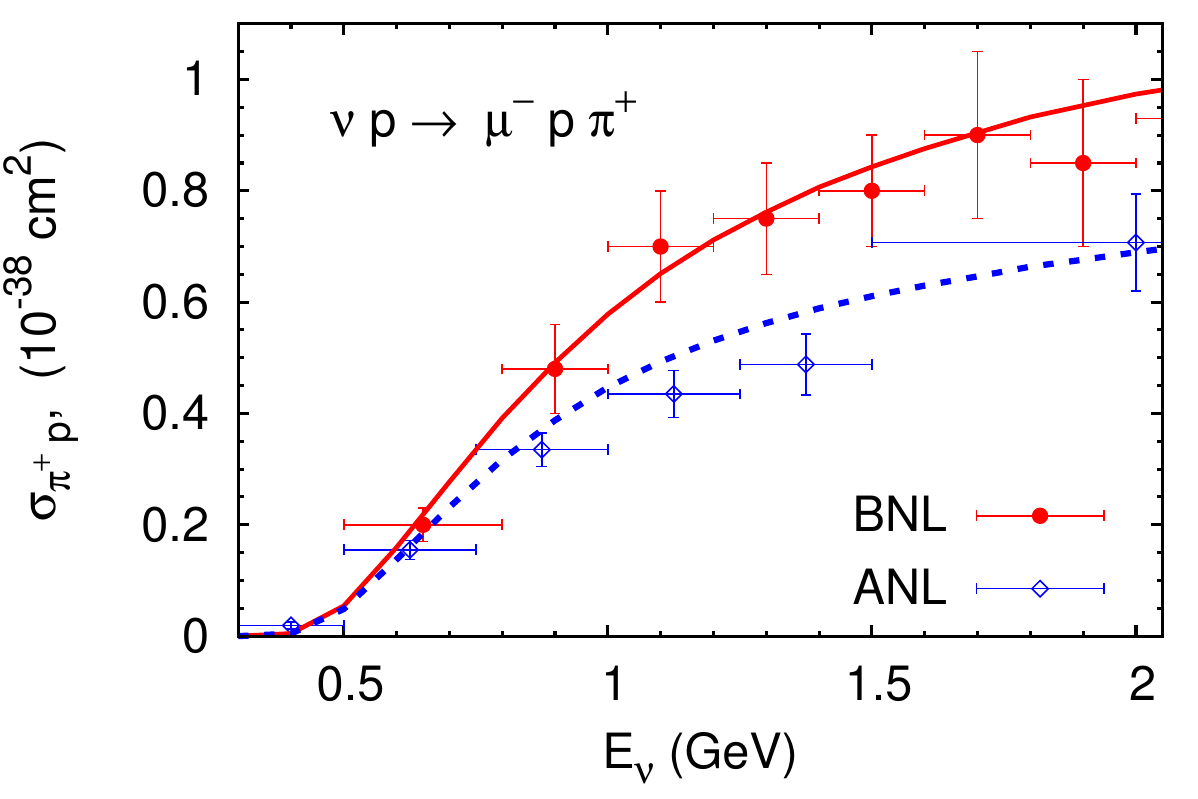}
\end{minipage}
\hfill
\begin{minipage}[c]{0.32\textwidth}
\includegraphics[width=\textwidth]{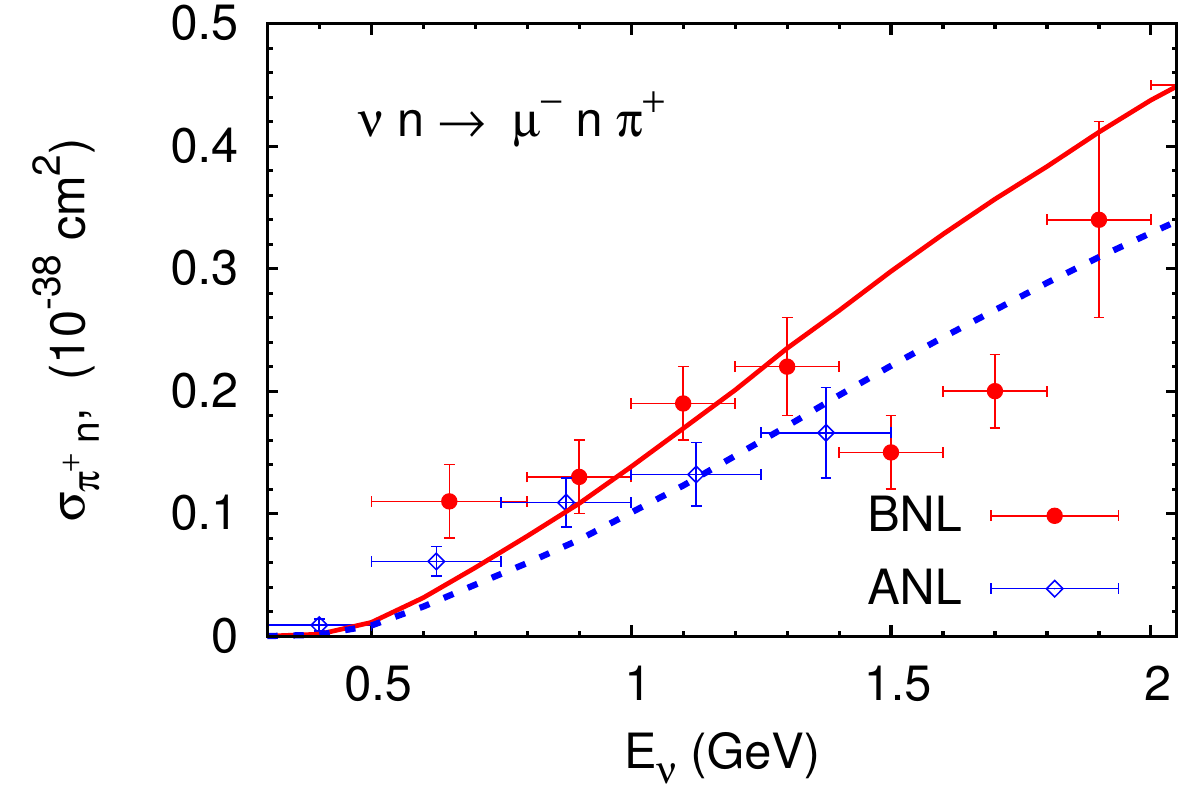}
\end{minipage}
\hfill
\begin{minipage}[c]{0.32\textwidth}
\includegraphics[width=\textwidth]{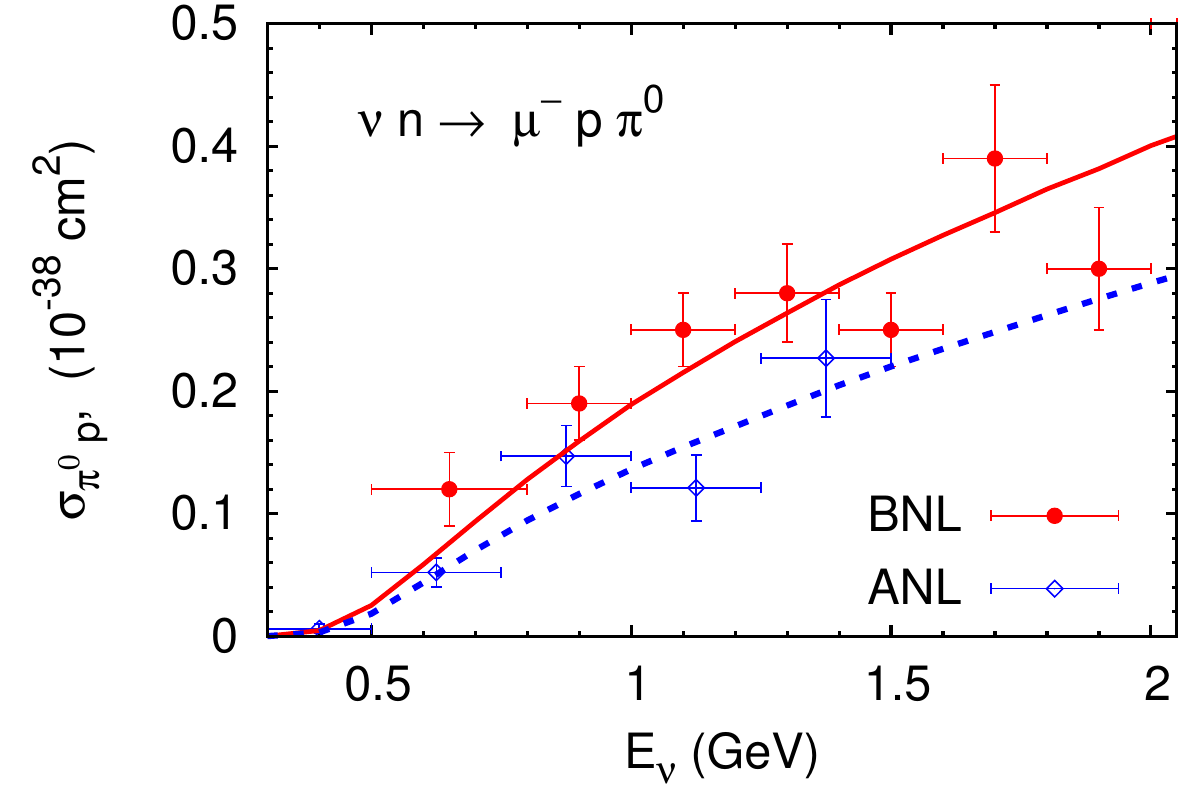}
\end{minipage}
\caption{Single-pion production cross section on proton and neutron targets obtained
in the BNL \cite{Kitagaki:1986ct} (circles; solid lines)
and the ANL experiments \cite{Radecky:1981fn}  (diamonds; dashed lines).
The curves give the lower (ANL-tuned) and upper (BNL-tuned) boundaries on the elementary input as used in GiBUU.}
\label{fig:ANLBNL1pidata}
\end{figure}

A necessary input into all calculations of neutrino-nucleus cross sections are
the elementary reaction cross sections.
For the \onepion production the most relevant ones are the ANL~\cite{Barish:1978pj,Radecky:1981fn}
and BNL~\cite{Kitagaki:1986ct} data for scattering off hydrogen and deuterium. Until
our recent publication~\cite{Lalakulich:2012cj} we had used the ANL data as default
input in the GiBUU code, because for these the absolute values of the cross section
for $\dd\sigma/\dd Q^2$ were given.
In this talk we now also show results obtained by using the BNL pion production cross sections
as input in order to explore the consistency of both input data sets with
the recent MiniBooNE data and to explore the sensitivity of the nuclear data to the
elementary interactions.

Fig.~\ref{fig:ANLBNL1pidata} shows that the data obtained in the BNL
experiment are generally higher than those from the ANL experiment.
In order to obtain an estimate for the systematic uncertainty in our
calculations, we fit here both of these data sets separately.
The curves in Fig.~\ref{fig:ANLBNL1pidata} show the theoretical description, obtained from
these fits and used as input in GiBUU. We consider them as lower and upper bounds of
the elementary cross section.

\subsection{How medium modifications of the $\Delta$ resonance properties
		influence the calculated cross section}

The GiBUU  simulation code, which besides describing many other nuclear reactions can also be used as neutrino event generator
was already described at the talk at this conference. There we have also emphasized the importance
of final state interactions of outgoing hadrons during their propagation throughout the nucleus.
The verification was provided by comparison with the photoproduction data
for neutral pions. Here we use the  same set of data to demonstrate the importance of taking into account
the medium modification of $\Delta$ resonance properties.
		
In nuclear medium the width $\Gamma$ of any baryon acquires a collisional contribution
\begin{equation}
\Gamma_{\rm coll} = - \frac{2}{\sqrt{p'^2}} \mathrm{Im} \Sigma_{\rm coll}(p'^2)  ~,
\end{equation}
where the selfenergy $\Sigma_{\rm coll}$ is a function of density and momentum.
For the $\Delta$ resonance the model for $\Sigma_{\rm coll}$ was proposed
by Oset and Salcedo (OS) in \cite{Oset:1987re}.
All calculations in this paper, except otherwise noted, have been done using this collisional width.

\begin{figure}
\includegraphics[width=\columnwidth]{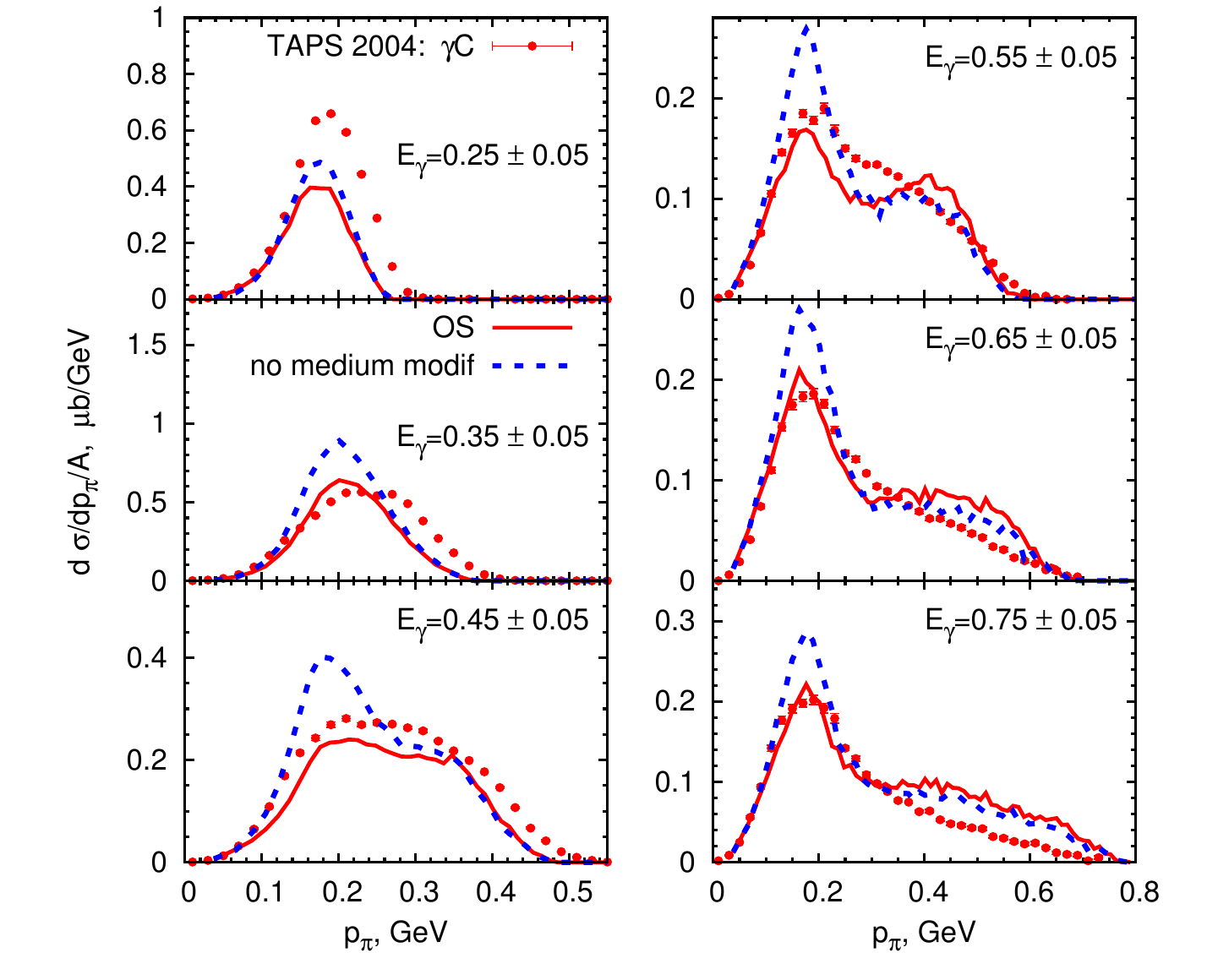}
\caption{Momentum distributions of the outgoing pions for inclusive $\pi^0$ production in scattering
 of photons of energies $0.2-0.8\GeV$ carbon.}
\label{fig:photoprod-oset-false}
\end{figure}

Fig.~\ref{fig:photoprod-oset-false} compares the photoproduction data with the results
of GiBUU calculations obtained with and without OS medium modifications.
Calculations using the OS model (red solid lines) are in general agreement
with the  data~\cite{Krusche:2004uw} (red circles), while calculations
without medium modifications (blue short-dashed lines)
noticeably overestimate the cross section in the $\Delta$ peak region.

\begin{figure}[htb]
\includegraphics[width=0.9\columnwidth]{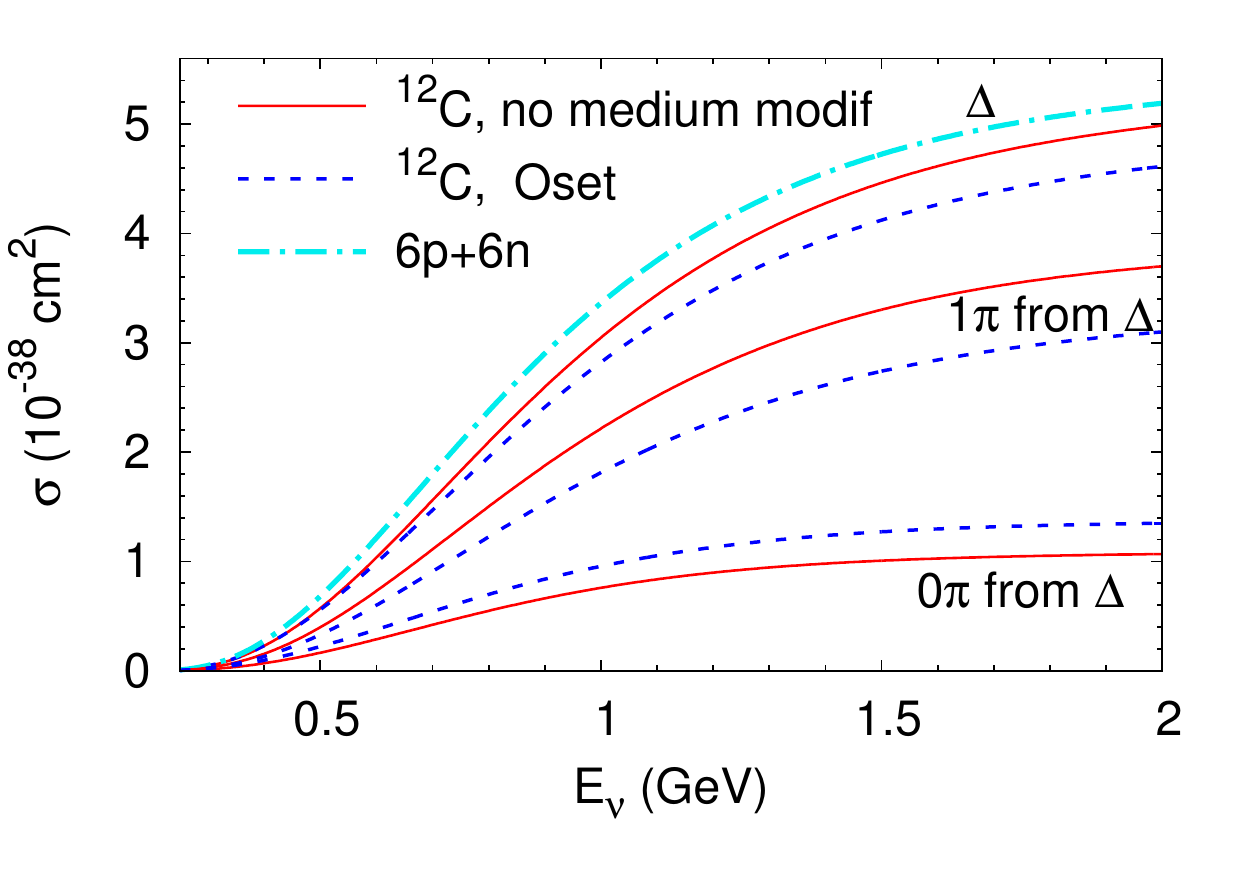}
\caption{Cross sections for $\Delta$ production, $1\pi$-production originating from $\Delta$ and
$0\pi$-production originating from $\Delta$  with and without the OS medium modification \cite{Oset:1987re}.}
\label{fig:carbon-Delta-1pi-nopi}
\end{figure}

\begin{figure}[htb]
\includegraphics[width=0.9\columnwidth]{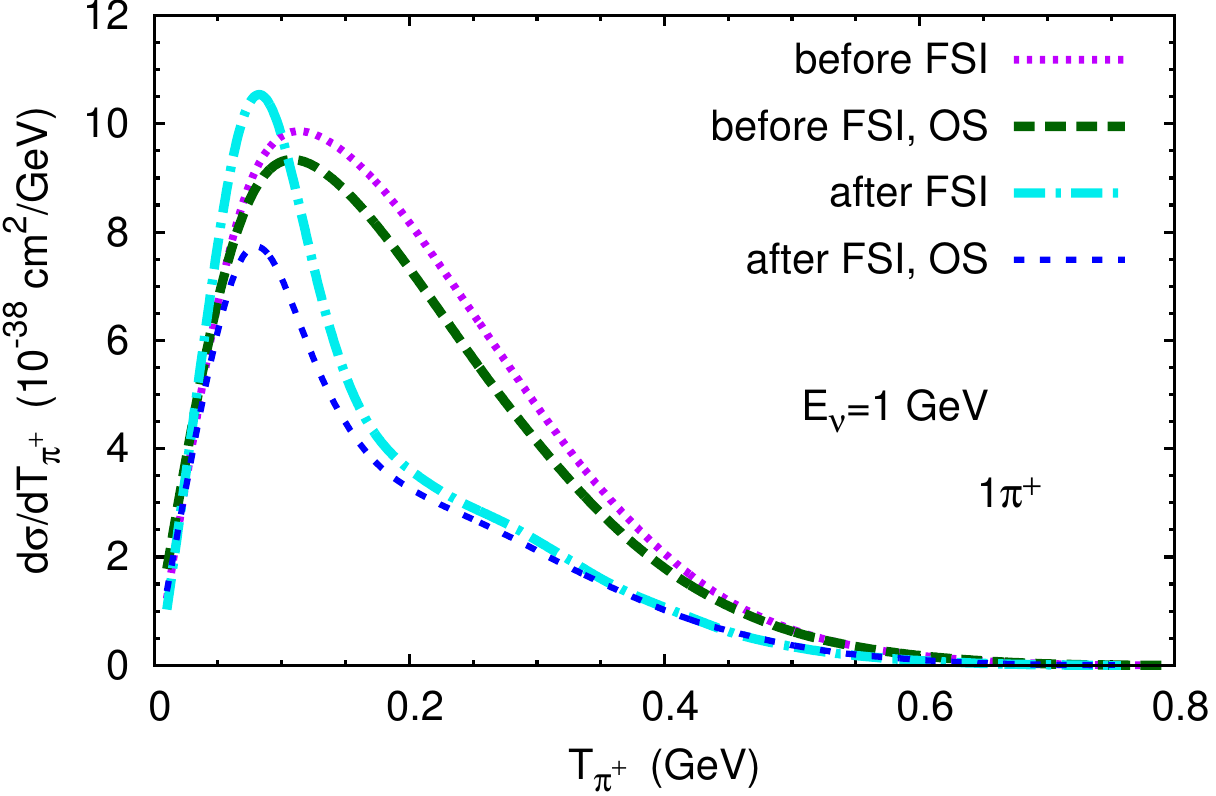}
\caption{Kinetic energy distribution of $\pi^+$ produced in neutrino scattering off carbon
 through the weak production of the $\Delta$ resonance and its following decay.
 The neutrino energy is $E_\nu=1\GeV$. The curves labeled OS were obtained using the OS in-medium
 collisional width of the $\Delta$.}
\label{fig:carbon-kinetic}
\end{figure}

The influence of $\Delta$ medium modifications on neutrino reactions is illustrated
in Figs.~\ref{fig:carbon-Delta-1pi-nopi}, \ref{fig:carbon-kinetic}.
The elementary input is the cross section for 6 free neutrons
and 6 free protons, all at rest (cyan dash-dotted line labeled ``6p+6n''). The nuclear effects,
namely the Fermi motion, the binding nuclear potential and the Pauli blocking of the outgoing proton,
decrease the cross section by around $5\%$ (red solid lines labeled ``${}^{12}$C'').

Taking into account the OS modification of Delta properties (blued dashed curve labeled ``${}^{12}$C, OS'')
leads to an additional decrease  of the $\Delta$-production cross section by another $5-8\%$.
This decrease is a consequence of the shift of strength from the $\Delta$ peak position
to larger masses due to the increased broadening. This extra strength at the higher masses is,
however, cutoff by the form factors so that the net effect is a lowering of the cross section.

A similar decrease, as expected, is observed  for the \onepion cross section, with the effect being
even larger: the cross section is decreased by $15-20\%$.
This additional suppression is due to the fact that the $\Delta$ now has an increased
collision width so that it can undergo $\Delta + N \to N N$ before the pion decay has taken place.
The same process, on the other hand, increases the cross section for events with no pions in the final state.
We stress that it is essential to treat the collisional broadening of the $\Delta$ spectral function consistently with
the $\Delta N$ collisions.


Fig.~\ref{fig:carbon-kinetic} shows the kinetic energy energy distribution of the $\pi^+$ originating from
$\Delta$ decay. Before FSI the distribution has a rather broad peak at $T\approx 0.1 \GeV$,
which is decreased by about $5\%$, if the OS modification of the $\Delta$ properties is taken into account.

After FSI, the OS modification of the $\Delta$ properties decreases the pion distribution significantly more,
by more than $20\%$ at the peak. This is due to the fact that together with the collisional broadening of
the $\Delta$ resonance explicit 2-body and 3-body collision terms are now active.
Thus, before having a chance to decay to a pion a significant part of $\Delta$s is absorbed in the nucleus.
The overall effect is a decreased pion production cross section which increases the number of pionless
events and thus serves as a source of the fake QE-like events.
It is worthwhile to point out already here that the special shape in the pion momentum distribution after
FSI is not affected by the OS modification.

\section{Output}

In the following discussions the data from~\cite{AguilarArevalo:2010bm,AguilarArevalo:2010xt}
are plotted versus reconstructed neutrino energy, while the theoretical curves presented
here are versus real energies.
We have found that the energy reconstruction method used in the experiment
\cite{AguilarArevalo:2010xt}, that relies only on the kinematics of the outgoing muon
and pion, is quite reliable, so we neglect the difference.
For the later comparisons it is also essential to note
that all the experimental cross sections for $\pi^+$ production were obtained with
the full MiniBooNE flux. Thus, for positively charged pions the distributions with
respect to muon or pion energy do not depend on any energy reconstruction scheme.
This is not so for the $\pi^0$ data where a cut for the (reconstructed) neutrino
energies was imposed: only neutrino energies between 0.5 and 2.0 GeV were taken
into account \cite{AguilarArevalo:2010xt}. Since this cut is not possible without a
generator-based reconstruction procedure the $\pi^0$ data may thus contain some model dependence.
We also note that all our earlier calculations published in
Refs.\ \cite{Leitner:2008wx,Leitner:2009de,Leitner:2009ec,Lalakulich:2011ne} used the full
MiniBooNE flux, without this cut, for all charge states.
The net result of using now the neutrino energy-window for $\pi^0$ is an increase of the
calculated flux-averaged cross sections for neutral pions compared to the earlier calculations.

\subsection{Inclusive cross section and lepton observables}

\begin{figure*}[!hbt]
\includegraphics[width=\textwidth]{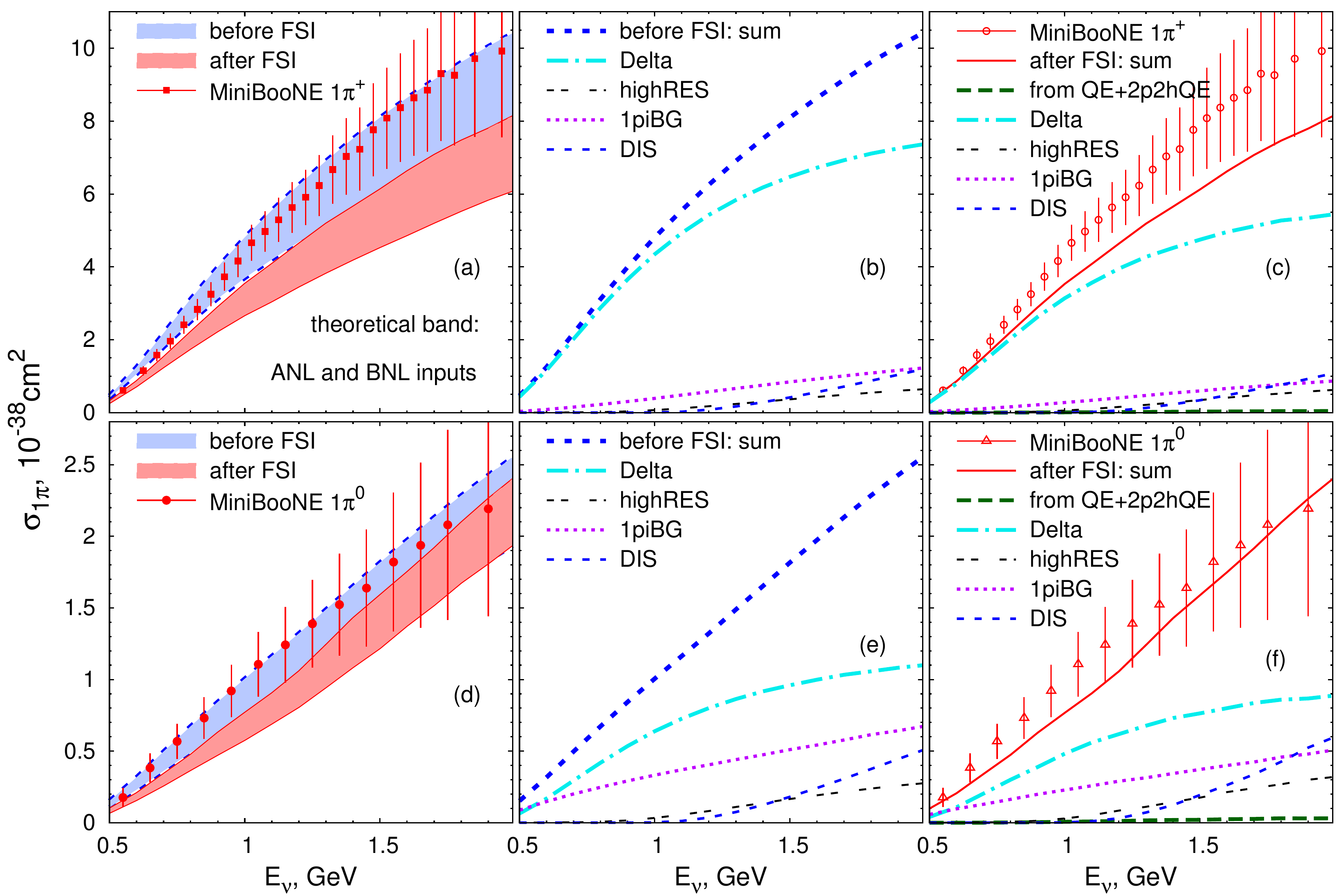}
\caption{(Color online) Integrated cross section for  1$\pi^+$ (a,b,c) and
1$\pi^0$(d,e,f) CC production versus neutrino energy. Panels (a) and (d) contain
the results of calculations using ANL- (lower curves) and the BNL-tuned (upper curves)
elementary input both before and after FSI.
Panels (b) and (e) show the contributions of various event origins to the calculated
cross sections before FSI.
Panels (b) and (f) show the same after FSI. Here the BNL-tuned input has been used.
Data are from~\cite{AguilarArevalo:2010bm,AguilarArevalo:2010xt}.}
\label{fig:MB-lepton-Enu-QEDelta}
\end{figure*}

The cross sections for \onepion production versus neutrino energy for CC $1\pi^+$ and $1\pi^0$ production
are  shown in Fig.~\ref{fig:MB-lepton-Enu-QEDelta}.

Figs.~\ref{fig:MB-lepton-Enu-QEDelta}a,b,c show the results for $1\pi^+$ production.
As expected, for a nuclear target (CH$_2$ in this case) the BNL elementary input leads to a
significantly larger cross section than the
ANL input, simply reflecting that fact that already the elementary BNL cross sections
are about 30\% larger than those from the ANL experiment (see Fig.\ref{fig:ANLBNL1pidata}).
For both inputs the calculated cross section after FSI (red solid lines)
is about 30\% lower than the
corresponding cross section before FSI (blue short-dashed lines).

Before FSI, the most of $1\pi^+$ events
come from initial  $\Delta$ resonance production and its following decay (cyan dash-dotted line).
This channel is dominant up to about 0.9 GeV incoming neutrino energy.
Some  events are background ones (dotted line in Fig.~\ref{fig:MB-lepton-Enu-QEDelta}b).
There is also a small contribution from higher resonances,
and at $E_\nu> 1.2 \GeV$ DIS processes start to play an increasing role.

FSI noticeably decrease the $\Delta$-originated \onepion production
(as comparison of panels (b) and (c) shows)
due to the absorption $N\Delta\to NN$; the similar process is possible also
for other resonances. Once a pion is produced, independent of its origin, it may also undergo
a charge-exchange $\pi^+ n \to \pi^0 p$ process, which depletes the $\pi^+$
channel as the dominant one.
The ensuing reduction of the $1\pi$-background channel due to FSI  is also noticeable.
Other possibilities for pions to disappear include $\pi N \to \omega N$,
$\phi N$, $\Sigma K$, $\Lambda K$.

A minor amount of pions comes from the initial QE vertex (green long-dashed line),
which is only possible due to FSI, when the outgoing proton is rescattered.
Here the main contribution is from  the $p N \to N' \Delta \to N' N^{''} \pi$ reaction.
Other possibilities to create pion during the FSI would be $\omega N \to \pi N$,
$\phi N \to \pi N$, $\pi N \to \pi \pi N$.

For the $1\pi^0$ production in Figs.~\ref{fig:MB-lepton-Enu-QEDelta} d,e,f
the curves before and after FSI are not so different. This is mainly due to the fact, that
side feeding from $\pi^+ n \to \pi^0 p$, which decreases the charged pion output, simultaneously
increases the neutral pion output.
The reverse process gives only minor relative contribution, because the initial $\pi^0$ production cross section is around 5 times lower.
This is, however, partly compensated by charge exchange to the $\pi^-$ channel through
$\pi^0 n \to \pi^- p$ and other channels mentioned above for charged pion production.

For the $1\pi^0$, the $\Delta$ channel is less important than in the case of charged pions,
while background processes, and at higher energies higher resonances  and DIS play a
relatively larger role than for $\pi^+$.
These two latter processes ensure the change in slope at about 1.2 GeV,
which reduces the  the discrepancy between the theoretical curves and data.

\begin{figure*}[!hbt]
\begin{minipage}[c]{0.48\textwidth}
\includegraphics[width=\textwidth]{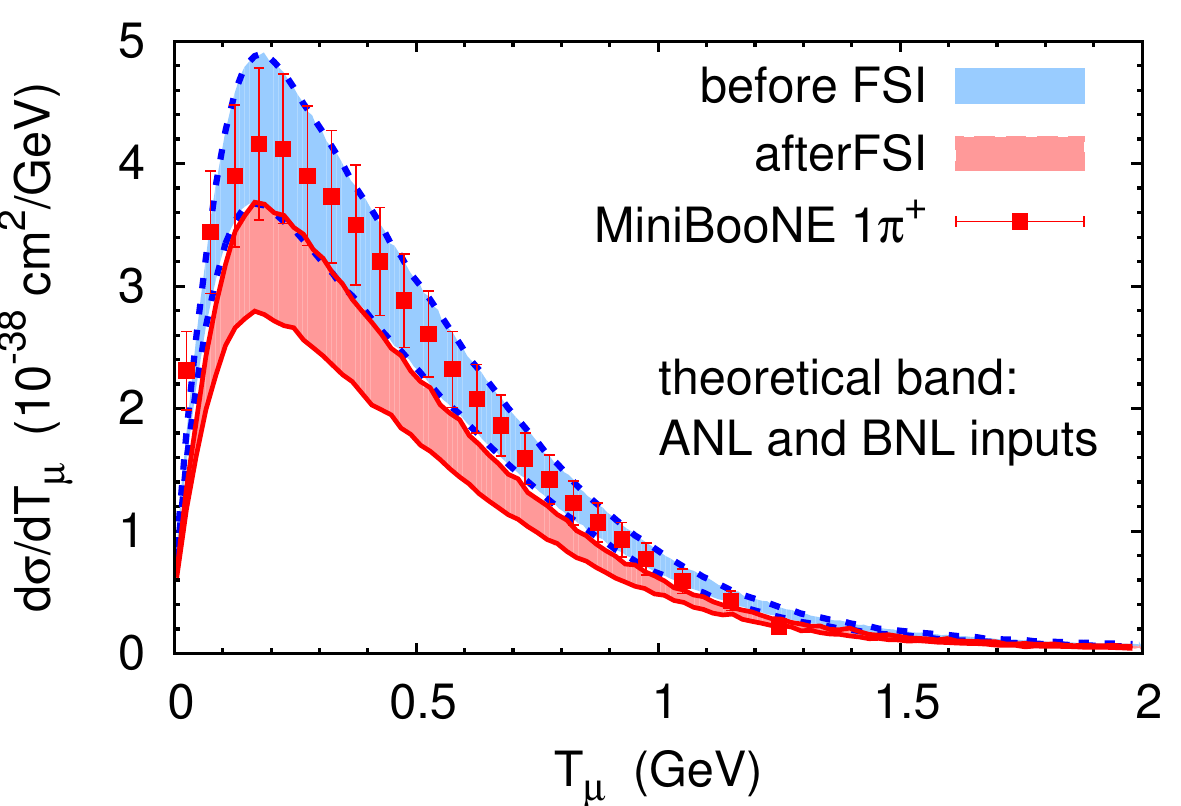}
\end{minipage}
\hfill
\begin{minipage}[c]{0.48\textwidth}
\includegraphics[,width=\textwidth]{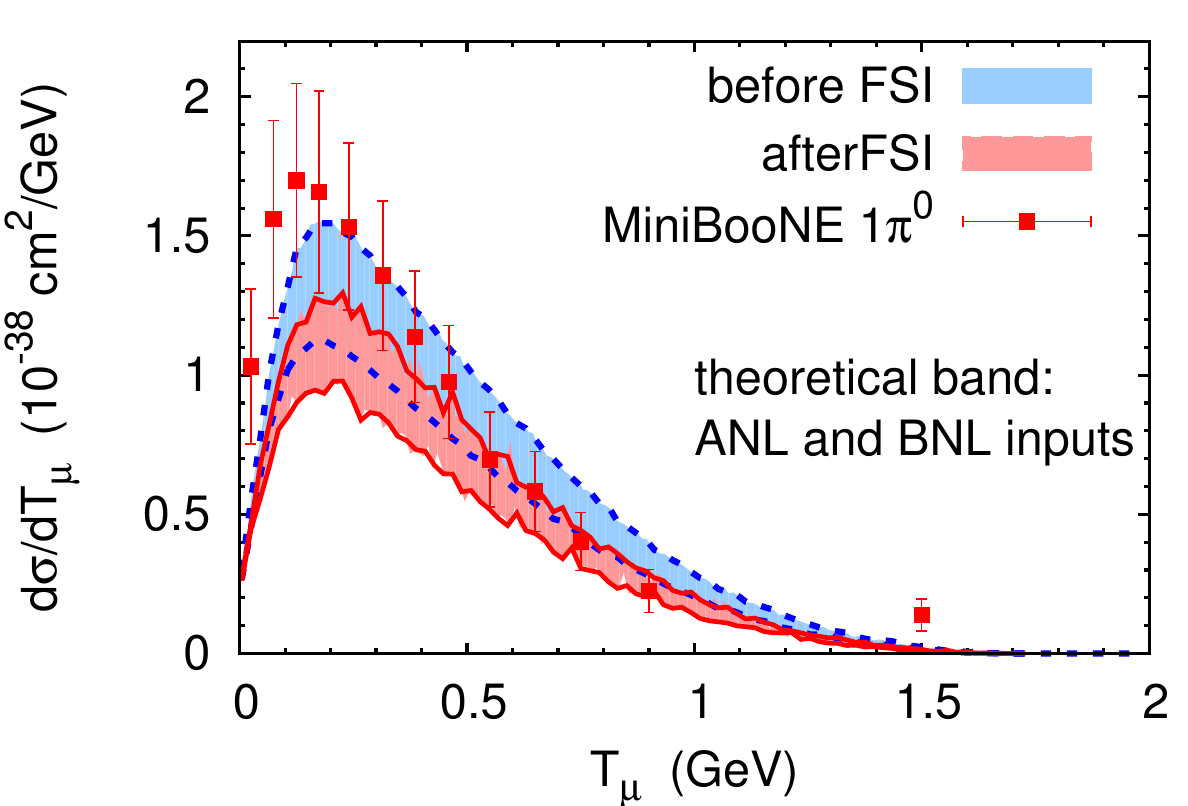}
\end{minipage}
\caption{(Color online) Kinetic energy distributions of the outgoing muons
for the  1$\pi^+$ and 1$\pi^0$ production in the MiniBooNE.
Data are from~\cite{AguilarArevalo:2010bm,AguilarArevalo:2010xt}.
The blue dashed curves give the results before FSI,
the red solid curves those with all FSI included.}
\label{fig:MB-lepton-Ekin}
\end{figure*}

The MiniBooNE collaboration has recently also published distributions versus muon kinetic energy,
$d\sigma/dT_{\mu}$, muon angle with respect
to the neutrino  beam direction, $d\sigma/d\cos\theta_{\mu}$, and squared four-momentum transfer,
$d\sigma/dQ^2$~\cite{AguilarArevalo:2010bm}. All of them are calculated within the GiBUU and
compared to the data in Ref.~\cite{Lalakulich:2012cj}.
Here we present only the kinetic energy distributions in Fig.~\ref{fig:MB-lepton-Ekin}.

It shows that FSI hardly influence the shape of the distributions
versus muon kinetic energy, the same is true for other muon observables.
This insensitivity is due to the fact that the only effect of FSI on these muon observables is that
they can remove events in which an initially produced pion (or $\Delta$) was later on reabsorbed
and bring in events in which the pion was produced only during FSI.

For $1\pi^+$ production, when referring to the shape-only comparison,
the calculated curves correspond to the data. However, the muon kinetic energy distribution is consistently 20\%
(upper boundary) to 60\% (lower boundary) lower than the data, see Fig.~\ref{fig:MB-lepton-Ekin}.
For $1\pi^0$ production, the shape of the curves evidently differs from the shape of the data.

\subsection{Pion Observables}

\begin{figure*}[!hbt]
\begin{minipage}[c]{0.48\textwidth}
\includegraphics[width=\textwidth]{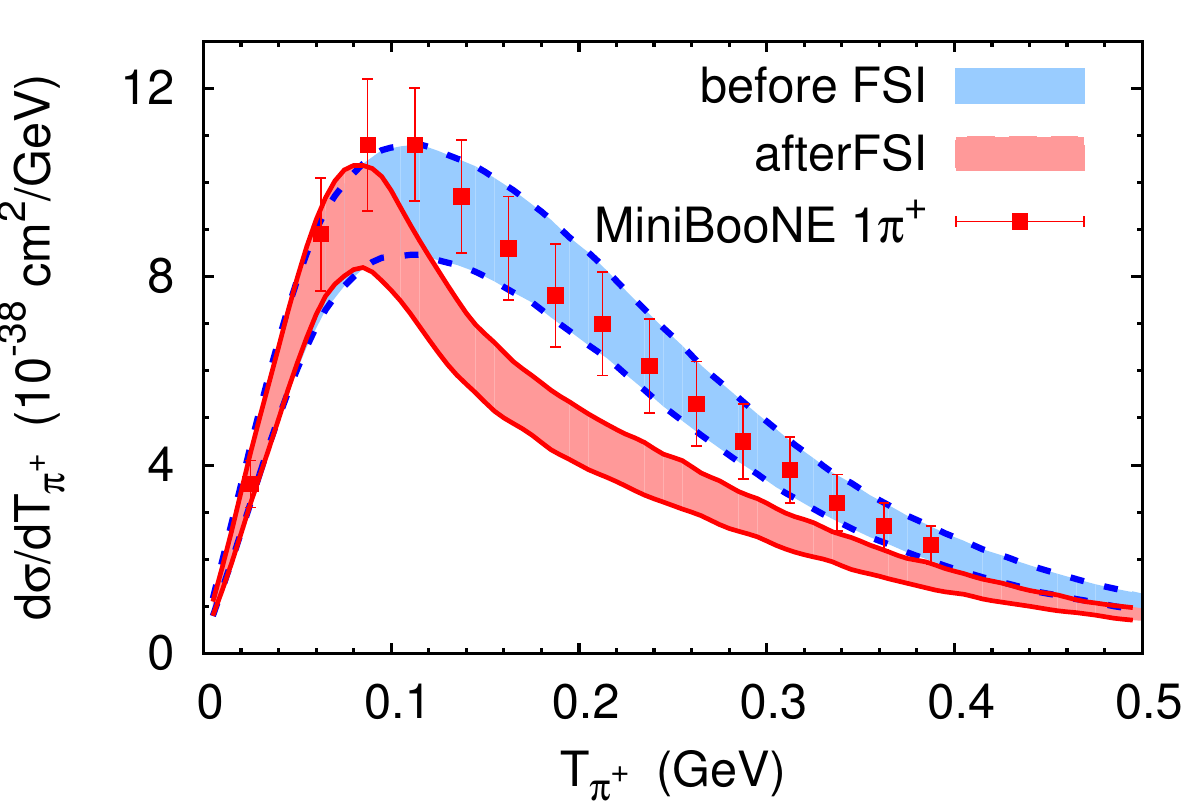}
\end{minipage}
\hfill
\begin{minipage}[c]{0.48\textwidth}
\includegraphics[width=\textwidth]{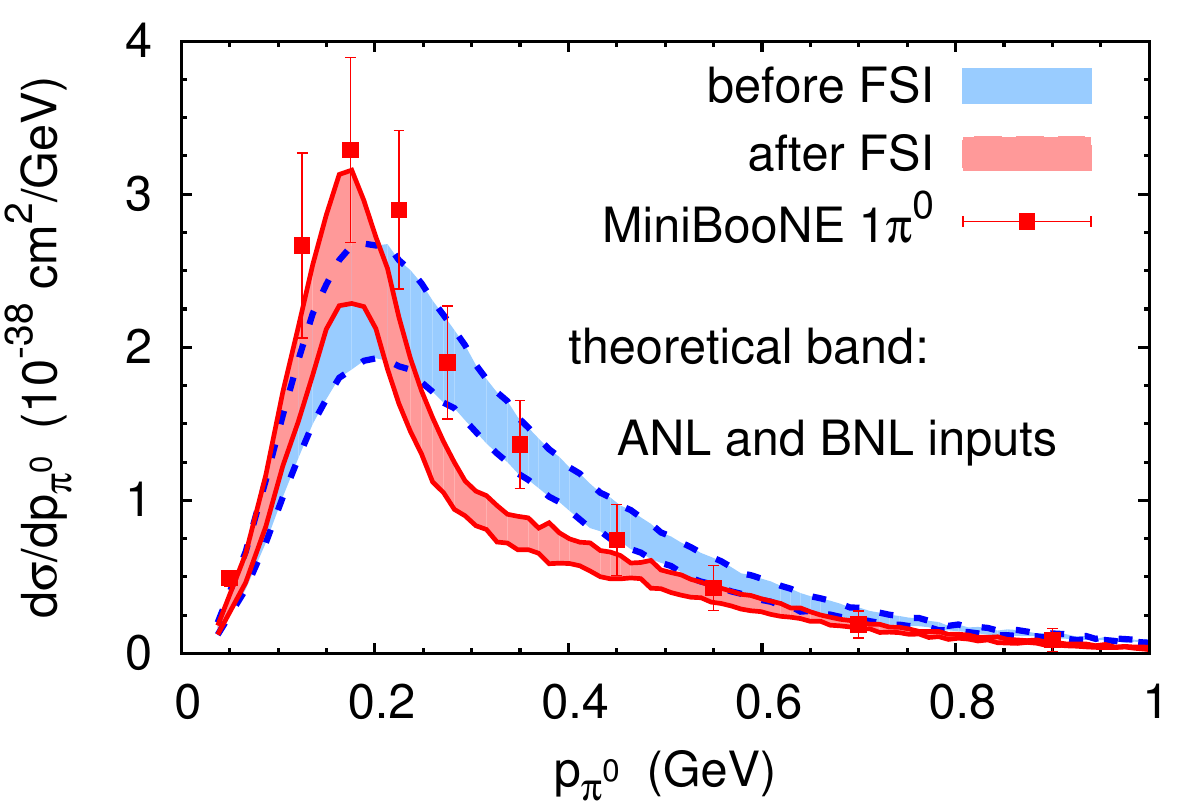}
\end{minipage}
\caption{(Color online) Kinetic energy distribution of the outgoing $\pi^+$ and momentum distribution of
the outgoing $\pi^0$ for \onepion production at MiniBooNE.
Data are from~\cite{AguilarArevalo:2010bm,AguilarArevalo:2010xt}.
The curves are as in Fig.\ \ref{fig:MB-lepton-Ekin}.}
\label{fig:MB-pion-dTkin}
\end{figure*}

\begin{figure*}[!hbt]
\begin{minipage}[c]{0.48\textwidth}
\includegraphics[width=\textwidth]{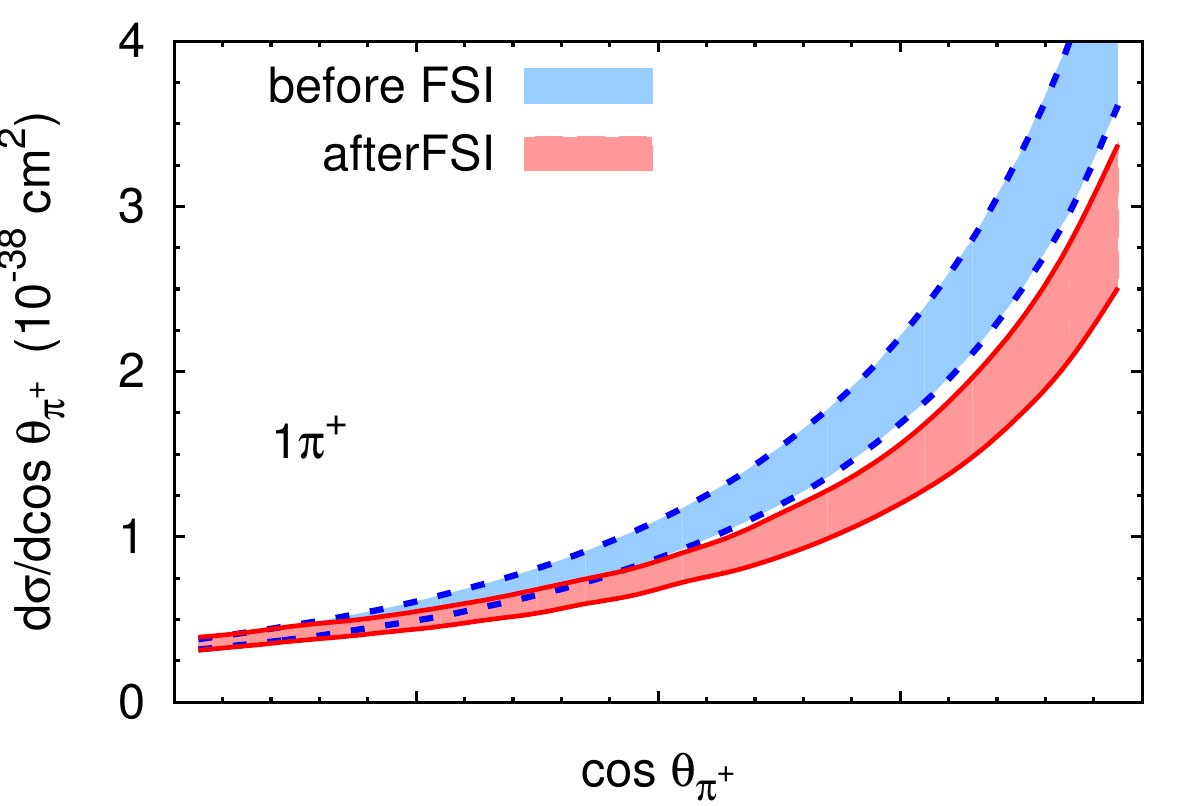}
\end{minipage}
\hfill
\begin{minipage}[c]{0.48\textwidth}
\includegraphics[width=\textwidth]{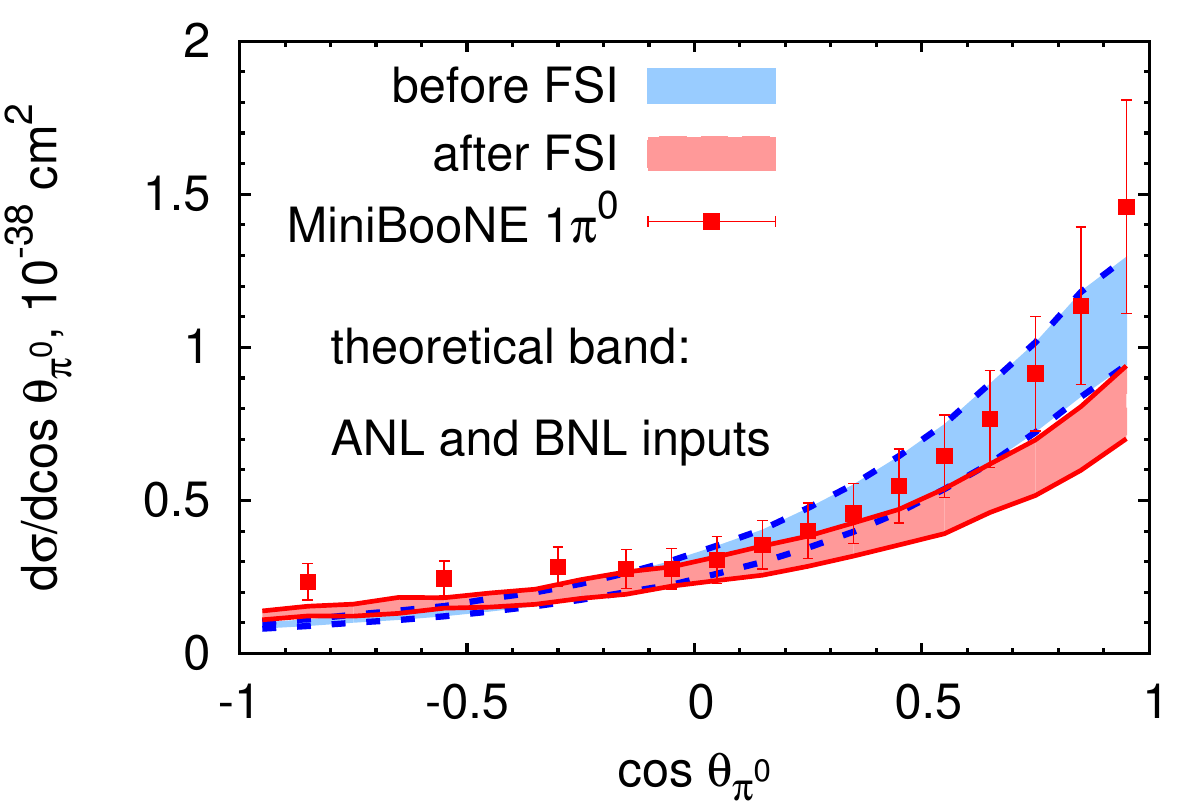}
\end{minipage}
\caption{(Color online) Distribution of the outgoing $\pi^+$  and $\pi^0$ in their angle relative to neutrino beam  for \onepion production at MiniBooNE.
Data are from~\cite{AguilarArevalo:2010xt}.
The curves are as in Fig.\ \ref{fig:MB-lepton-Ekin}.}
\label{fig:MB-pion-dcospion}
\end{figure*}

Fig.~\ref{fig:MB-pion-dTkin} presents the results of our calculations for the kinetic
energy distribution of the outgoing pions. Unlike the case of muon-related observables,
for pion-related distributions the FSI noticeably change  their shape.
Surprisingly, the shape before FSI is similar to that observed  in the data.
Nevertheless, the final state interactions must be inevitable for nuclear targets.
After FSI, however, the shape of the calculated distributions looks markedly different.
In particular, there is a significant lowering around $T_\pi \approx$ 0.2 GeV for $\pi^+$ and at
around $p_{\pi}^0 \approx 0.3$ 0.3 GeV, as a direct consequence of the $\Delta\pi N$ dynamics in nuclei.

The following processes are important for this structure: pion elastic scattering in the FSI decreases
the pion energy, thus depleting spectra at higher energies and accumulating strength at lower energies.
Simultaneously, there is charge exchange scattering. At the same time pions are mainly absorbed via the
$\Delta$ resonance, that is through $\pi N \to \Delta$ followed by $\Delta N \to N N$, which leads to the
reduction in the region of pion kinetic energy $0.1-0.3 \GeV$. For $\pi^0$ production the additional
increase of the cross section at lower energies comes from the side feeding of the $\pi^0$ channel from the
dominant $\pi^+$ channel due to the charge exchange scattering $\pi^+  n \to \pi^0 p$.

Inverse feeding
$\pi^0 p \to \pi^+ n$ is suppressed, because at the energies under consideration, about 5 times less
$\pi^0$s than $\pi^+$s are produced. The change of the shape of the spectra due to FSI is similar to
that calculated for neutral current 1$\pi^0$ production in \cite{Leitner:2008wx,Leitner:2009de}.

As we already emphasized in another talk at this conference, the particular shape calculated
here for the neutrino-induced pions is in line with that observed
experimentally in $(\gamma, \pi^0)$ production on nuclear targets \cite{Krusche:2004uw}. It has also
been observed in the independent calculations of Hernandez et al.\ \cite{Hernandez:2012}.
Since the shape depends on FSI and since the FSI are the same
in both neutrino-induced and photon-induced reactions the absence of this special shape in the
neutrino data is surprising.

Fig.~\ref{fig:MB-pion-dcospion} shows the pion angular distribution of the of the CC $1\pi^+$ and $1\pi^0$
production.  Here data are available for $\pi^0$ only; for forward scattering
they are significantly higher than the lower boundary of our calculations,
and only slightly higher compared to the upper boundary. The forward peaking is reproduced,
but it is not as strong as exhibited by the data.

\section{Conclusions}
In conclusion we summarize here a few essential points.

\begin{itemize}
 \item Any neutrino event generator should have all initial reaction mechanisms (quasi-elastic scattering,
 $\Delta$ production, production of higher resonances, background 1- and many-pion production, DIS) under
 control. Even the MiniBooNE data require the inclusion of processes beyond the $\Delta$ excitation.

 \item There is a strong dependence of theoretical results on in-medium Delta properties.

\end{itemize}

Finally, we stress that because of the strong dependence of the results on the elementary pion production cross section new, improved data for neutrino-induced pion production on elementary targets are clearly needed to resolve the ambiguity in pion production on nuclear targets. Only when such data have become available the question of a better determination of the axial coupling to nucleon resonances can be approached again.

This work is supported by DFG and BMBF.

\bibliographystyle{aipproc}
\bibliography{nuclear}

\end{document}